\documentclass[notoc]{JHEP} 
\usepackage{amsmath}
\usepackage{epsfig}
\usepackage{amssymb,amsfonts}

\def\be{\begin{equation}}
\def\ee{\end{equation}}
\def\ba{\begin{eqnarray}}
\def\ea{\end{eqnarray}}

\newcommand{\C}{{\cal C}}

\def\sqr#1#2{{
\vcenter{\vbox{\hrule height.#2pt
\hbox{\vrule width.#2pt height#1pt \kern#1pt
\vrule width.#2pt}
\hrule height.#2pt}}}}
\boldmath
\title{Gravitational couplings of orientifold planes}
\unboldmath
\author{Pierre Henry-Labord\`ere  and Bernard Julia\\
Laboratoire de Physique Th{\'e}orique
de l'Ecole Normale Sup{\'e}rieure\thanks{Unit{\'e} mixte  du
CNRS et de l'Ecole Normale Sup{\'e}rieure,  
UMR 8549.}\\ 
24  rue Lhomond, 75231 Paris Cedex 05, France\\
\email{phenry@lpt.ens.fr}\\
\email{bernard.julia@lpt.ens.fr}}

\abstract{
We reanalyse the gravitational couplings of the
perturbative orientifold planes $Op^-$, $Op^+$ (and D-branes).
We first compute their $D_{-1}$ instantonic corrections for $p=3$. Then,
by using U-dualities, we obtain the Wess-Zumino terms of orientifolds with
RR flux for $p \leq 5$. The expressions for the effective actions can be   
partially checked via M-theory. We point out a previous oversimplification
and we show in fact that the difficulty still stands in the way of the
full computation of 7 Brane instanton corrections.
}
\preprint{LPT-ENS/01-38, hep-th/0112nnn}

\begin{document}

                                                                                      \section{Introduction}

Four kinds of orientifold p-planes with sixteen supersymmetry charges
exist: $Op^-$, $Op^+$ and
for ($p \leq 5$) 
$\widetilde{Op^-}$ or $\widetilde{Op^+}$ 
\cite{wit1,han}.
Only $Op^-$ and $Op^+$ can be
described perturbatively (in the string coupling)
 as the fixed planes of the orientifold
projection $\Omega$ which acts on the world-sheet and on the
Chan-Paton factors. $\widetilde{Op^-}$ can be interpreted as the bound
state of an $Op^-$ and half a $Dp$ brane, as one obtains the
right RR charge for ${1 \over 2}$ $Dp$ on top of an $Op^-$. Then adding
$n$ more $Dp$'s, we obtain the gauge group $SO(2n+1)$
which extends the $SO(2n)$ realization.
The transverse space to the orientifolds contains a $RP^2$
(resp. $RP^{5-p}$)
cycle  through which the fluxes 
$\vartheta_{NS}=\int_{RP^2} B_{NS}$ and $\vartheta_{R}=\int_{RP^{5-p}}
C_{5-p}$ are 0 
or 1 mod two (Tab. \ref{ori}).

The effective world-volume actions of the $Op^-$, $Op^+$ planes and $Dp$-branes
can be determined perturbatively by a string computation \cite{cra,mor,ste} 
or by requiring
the consistent cancellation of anomalies that arise when pairs of
branes intersect \cite{gre,scr}. The perturbative formulae for the orientifold planes and the
$Dp$-branes are:
\begin{eqnarray}
WZ(Op^{-})& = & -2^{p-5}\int_{M_{p+1}} C \wedge \sqrt{\hat{L}(R/4)}
\label{Op^{-}} \\
WZ(Op^{+})& = &  2^{p-5}\int_{M_{p+1}} C \wedge
(2\sqrt{\hat{A}(R)}-\sqrt{\hat{L}(R/4)})
\label{Op^{+}}\\
WZ(Dp) & = &         \int_{M_{p+1}} C \wedge ch(E) \wedge \sqrt{\hat{A}(R)}
\label{Dp}
\end{eqnarray}
where $C$ is the sum of RR forms $\sum_{i}C_i$.
$\hat{A}(R)$  and $\hat{L}(R)$ are respectively  the Dirac roof genus
and the Hirzebruch polynomial given in terms of Pontryagin classes by:
\begin{eqnarray}
\hat{A}(R)& = & 1-\frac{1}{24}p_1+\frac{1}{5760}(7p_1^2-4p_2) \\
\hat{L}(R/4)&= & 1+\frac{1}{48}p_1-\frac{1}{11520}(p_1^2-7p_2)
\end{eqnarray}
The Chern Character of the D-brane bundle appears too and the maximal dimension
possible is $p+1=10$. Let us notice that the difference
\eqref{Op^{-}}-\eqref{Op^{+}} only depends on the square root of the roof
genus whereas the first topological term \eqref{Op^{-}} depends only on the 
square root of the signature. These perturbative results are strikingly simple. 
\\
As we know the transformation of $B_{NS}$ and $C_{2}$ under $SL(2,{ \mathbb{Z}})$,
we may  deduce the congruence subgroups ${\cal G}$ of $Sl(2,{ \mathbb{Z}})$ 
that preserve the
various orientifold 3-planes (Tab. \ref{ori}). It should be stressed that the charges
$(\vartheta_{NS}, \vartheta_{R})$ transform  cogradiently to the gauge
fields $(B_{NS},C_2)$ under ${\cal G}$.
\\
In \cite{muk}, the gravitational coupling 
of an $\widetilde{Op^-}$ orientifold was computed by adding the Wess-Zumino
 terms of an $Op^-$ orientifold and half a $Dp$ brane. The latter was assumed 
to be  one half the Wess-Zumino 
action of a full $Dp$ brane. This hypothesis  seems too strong because 
the $Dp \over 2$ brane has to be  non perturbative to make a
perturbative $Op^{-}$ into a non perturbative $\widetilde{Op^-}$.    

We will see in section two that this hypothesis is incorrect. We 
will compute  gravitational couplings for $\widetilde{Op^-}$ and 
$\widetilde{Op^+}$ orientifolds that differ from those of \cite{muk}. 
To obtain the Wess-Zumino actions of $\widetilde{Op^-}$ and $\widetilde{Op^+}$,
we will use the various S-dualities that exist for $p=3$.
T-dualities reduce the problem to $p=3$, one then uses non
perturbative gravitational couplings for $D3$, heterotic string and CHL string
on $T^6$ (in fact perturbative computation and $SL(2,{ \mathbb{Z}})$
invariance lead to the same result up to normalisation for $D3$ and
heterotic string on  $T^6$) and computes those of $O3^-$ and  $O3^+$.
One finally gets $\widetilde{O3^-}$ and $\widetilde{O3^+}$
planes  by applying respectively  $S$ and $ST$ transformation
to $O3^+$.
We recall that an orientifold $O(p+1)^-$ (resp. $O(p+1)^+$)
gives by T-duality along it two $Op^-$  ( resp. two $Op^+$) and an orientifold 
$\widetilde{O(p+1)^-}$  (resp. $\widetilde{O(p+1)^+}$) gives  
one $\widetilde{Op^-}$ (resp. $\widetilde{Op^+}$)
and one  $Op^-$  (resp. $Op^+$) \cite{keu}. Flux conservation
and dualities relate the $R-R$ charges of the orientifolds.
By a perturbative computation, we
know that $O3^+$ has  charge ${1 \over 4}$. As the $C_4$ form is 
invariant under $SL(2,\mathbb{Z})$, we obtain that the $C_4$-charges for 
$\widetilde{O3^-}$ and $\widetilde{O3^+}$ are both ${1 \over 4}$. Then
by using 
T-duality we obtain that the charge of  $\widetilde{Op^-}$
(resp. $\widetilde{Op^+}$) is $-2^{p-5}+{1 \over 2}$ 
(resp. $2^{p-5}$ ). 

In section three, we relate these results to semi-classical
computations in ${\cal M}$ theory. They rely crucially on the $M5$
brane , on the ``orientifold''  ${\cal 
OM}5$ and on the ( $x^{11}$ delocalized) lift of $\widetilde{O4^-}$ \cite{gim}.

\begin{table}[http]
\begin{center}
\begin{tabular}{|c|c|c|c|c|}
\hline
{\rm Notation} &($\vartheta_{NS}$, $\vartheta_{R}$)  &  {\rm charge} 
$Q_p$ & {\rm Group} & Self-duality $p=3$ ${\cal G}$
\\
\hline
$Op^-$ &(0,0) mod. 2 &  $-2^{p-5}+n$ & $SO(2n)$  & $SL(2,{ \mathbb{Z}})$ 
\\
\hline
$Op^+$ &(1,0) mod. 2 &  $2^{p-5}+n$ & $Sp(n)$ & $\Gamma_{0}(2)=\lbrace \left(
\begin{array}{ll}
1
& \star \cr
0 & 1 \cr
\end{array}
\right) 
 \text{mod 2} \rbrace $

 \\
\hline
 $\widetilde{Op^-}$ &(0,1) mod. 2  & $-2^{p-5}+{1 \over 2}+n$ & $SO(2n+1)$ &
 $\Gamma^{0}(2)=\lbrace \left(
\begin{array}{ll}
1
& 0 \cr
\star & 1 \cr
\end{array} 
\right)
 \text{mod 2}
\rbrace$
\\
\hline
$\widetilde{Op^+}$ &(1,1) mod. 2  & $2^{p-5}+n$ & $Sp(n)$  & 
$\Gamma(2)=\lbrace \left(
\begin{array}{ll}
1
& 0 \cr
0 & 1 \cr
\end{array}
\right) 
\text{or}
\left(
\begin{array}{ll}
0
& 1 \cr
1 & 0 \cr
\end{array}
\right)
 \text{mod  2}
\rbrace$\\
\hline
 $n$ $Dp$& & $n$ & $U(n)$  & $SL(2,{ \mathbb{Z}})$
\\
\hline
\end{tabular}
\end{center}

\label{ori}
\caption{Orientifold $p$-planes with  $n$ $Dp$ branes. $\star$ stands for 0 or 1.}

\end{table}

\section{Dualities}

The Wess-Zumino actions written above can receive nonperturbative corrections 
coming from $D_{-1} $ instantons stuck on the
odd-planes. As the force between 
$D_{-1} $ and $D_{p} $ branes is  not repulsive 
 for p=3 and p=7,  3-planes and 7-planes may receive corrections. Moreover, 
only the terms
with at least $\frac{{\cal N}}{4}$ derivatives receive D-instanton 
corrections in the theories with ${\cal N}$ SUSY.
Indeed, a term with
$\frac{n_{f}}{2}$ derivatives  is transformed by supersymmetry in a
vertex with at most $n_{f}$ fermions.
As the $D_{-1} $ instanton leaves half of the ${\cal
N}$ SUSY unbroken, we have  $\frac{{\cal N}}{2}$ fermionic modes. 
To obtain a result different from zero when we integrate over the 
fermionic coordinates 
 we have the semi-classical rule:
$n_{f} \geq \frac{{\cal N}}{2}$. In the following section, we will consider
 theories with ${\cal N}=16$. The gravitational
couplings with at least 4 derivatives
 will receive D-instanton corrections. For example,
for the orientifold three-planes, only the ``BPS-saturated'' coupling 
 $\int C_0 p_1$ receives corrections from D-instantons.

We begin with  the $SO(32)$ heterotic string compactified on a two
torus of K\"ahler class $T_H=B_{NS}+iV$ and complex structure $U_H$ at
the $SO(8)^4$ point corresponding to two Wilson lines $Y_1=(0^4,0^4,{1 \over
2}^4, {1 \over 2}^4)$
and $Y_2=(0^4, {1 \over 2}^4, 0^4, {1 \over 2}^4)$. We first S-dualize
to obtain type I and then  apply a double T-duality along $T^2$. 
We obtain an orientifold from $IIB$ theory on 
${T^{2} / ({ \mathbb{Z}}_{2} (-)^{F_L} \Omega})$ with 16 D7's and 4 $O7^-$'s
where $\mathbb{Z}_{2}$ acts  as inversion on the two torus $T^2$,
$F_L$ is the left moving fermion number and $\Omega$ is the worldsheet
parity operator.
The moduli
 $\tau=C_0+ie^{-\phi}$ corresponding
to the $D_{-1}$ action
and $U_{IIB}$ are identified respectively with $T_H$ and $U_H$.
This is F-theory  on a special $K3$ surface (an  elliptic
fibration with base $ T^2 / { \mathbb{Z}}_2$ \cite{sen}).
 Each of the four singular fibers
corresponds to 4 $D7$ plus one orientifold $O7^-$
composed  of a (1,1) 7-plane and a (1,-1) 7-plane.
We recall that a $(p,q)$ 7-plane is a 7-brane on which a (p,q) string
can end \cite{sch}. We  prove in appendix B, that the action of an
$O7^-$ orientifold cannot be obtained by adding the Wess-Zumino
terms of a (1,1) and (1,-1) 7-planes. This is an important obstacle to 
deriving effective actions, we shall circumvent it by using S-duality in four 
dimensions.

Let us now compactify further the preceding configuration on a four 
torus $T^4$ and apply four  
T-dualities, we obtain type IIB theory compactified on a $T^6 /
\mathbb{Z}_{2}$\vspace{-.1mm}-\vspace{-.1mm}orientifold with 16 $D3$'s and 64
$O3^-$'s which  is  dual to the heterotic string
on $T^6$. We should also note that heterotic string theory
on $T^6$ is dual to type IIA theory compactified on $K3 \times
T^2$. The modulus $T_H$ is identified with the K\"ahler modulus $T_{IIA}$.
The term $\int f_{IIA}(T_{IIA})p_1$ in type IIA theory can    
be computed exactly from a one-loop calculation for this geometry
assuming no purely non perturbative cusp terms appear. 
Using the duality chain
we identify this term with the Wess-Zumino action for 16 $D3$ 
and 64 $O3^-$. Finally, we have 

\begin{equation}
\int \left(16f_{D3}(\tau)+64f_{03^-}(\tau)\right)p_1=\int f_{HET}(\tau)p_1=
\int \frac{6}{\pi}Re \left[ \, i \, \ln(\eta(\tau)) \right]p_1
\label{n1}
\end{equation}
where $\eta(\tau)$ is the famous Dedekind function and $Re$ is the
real part. The properties of
the theta functions that we
will use in this paper, have
been collected in  appendix A. The first term on the left handside is a sum because the two terms correspond to the same monodromy.

As a second theory, let us now consider a type  IIB orientifold with
 8 $D7$'s, 3 $O7^{-}$'s and one $O7^{+}$. It 
is  T-dual to type IIB on $T^2$ modded out by the worldsheet
projection $\Omega$ in the presence of a half-integral background flux of 
$B_{NS}$. This orientifold is equivalent to a compactification without
vector structure \cite{wit2,bia} and it is dual to the 8-dimensional CHL
string \cite{cha,lsch}.  By compactifying on  $T^4$ and applying again
four T-dualities along it
we obtain
that the type IIB orientifold  on $T^6 / \mathbb{Z}_2$
with 8 $D3$'s, 48 $O3^{-}$'s and 16 $O3^{+}$'s
is dual to the CHL string  on $T^6$. In \cite{greg},
the term $\int f_{CHL}(T_H)p_1$ has been computed exactly so we
obtain 
\begin{equation}
\int \left(8f_{D3}(\tau)+48f_{03^-}(\tau)+16f_{03^+}(\tau)\right)p_1=
\int f_{CHL}(\tau)p_1=
\int \frac{1}{\pi}Re \left[ \, i \, \ln(\vartheta_2(\tau)\eta^{3}(\tau)) \right]p_1 
\label{n2}
\end{equation}

Finally a IIB orientifold 
with 2  $O7^{-}$'s and 2 $O7
^{+}$'s is S-dual to an asymmetric ${ \mathbb{Z}}_2$ orbifold of IIB on $T^2$ where 
the ${ \mathbb{Z}}_2$ acts
as $(-)^{F_{L}}$ together with a translation on $T^2$.
There is  no more D-brane. We can again compactify on $T^4$ and T-dualize. We
obtain that IIB orientifold with 32 $O3^{-}$'s and 32 $O3
^{+}$'s is dual to an asymmetric ${ \mathbb{Z}}_2$ orbifold of IIB on $T^6$.
This model was argued to be U-dual to type IIA over $\frac{T^4}{\mathbb{Z}_2}
\times T^2$ compactification \cite{vafa}. 
The $\mathbb{Z}_2$ acts both as a twist on the
$T^4$ and as a shift on the two torus. The modulus $\tau$ is again
 identified with the K\"ahler modulus $T_{IIA}$. The coefficient of
$p_1$ has  been
computed exactly in type IIA. With the duality chains, we obtain that 
\begin{equation}
\int \left( 32f_{03^-}(\tau)+32f_{03^+}(\tau) \right)p_1=\int f_{IIB_{asym}}(\tau)p_1=
\int \frac{2}{\pi}Re \left[ \, i\, \ln(\vartheta_2(\tau)) \right]p_1
\label{n3}
\end{equation}

We now put together our three results: \eqref{n1}, \eqref{n2}, \eqref{n3},
and we obtain the following system 
for the gravitational couplings of the orientifold three-planes.
\begin{equation}
\left(
\begin{array}{lll}
16
& 64 &0  \cr
8 & 48 & 16 \cr
0 & 32 & 32 \cr
\end{array}
\right) 
\left(
\begin{array}{l}
f_{D3}(\tau)
 \cr
f_{O3^{-}}(\tau) \cr
 f_{O3^{+}}(\tau)\cr
\end{array}
\right) 
=
\left(
\begin{array}{l}
\frac{6}{\pi}Re \left[ \, i\, \ln(\eta(\tau)) \right]
 \cr
\frac{1}{\pi}Re \left[ \, i\, \ln(\vartheta_2(\tau)\eta^{3}(\tau)) \right] \cr
\frac{2}{\pi}Re \left[ \, i\, \ln(\vartheta_2(\tau)) \right] \cr
\end{array}
\right) 
 ,
\label{a}
\end{equation}
The system \eqref{a} is singular, in fact  the tadpole
cancellation condition for the
$C_4$ charges $Q_3$ leads to the constraint
$2f_{CHL}(\tau)=f_{HET}(\tau)+f_{IIB_{asym}}(\tau)$ which is satisfied by our 
expressions. This is a non trivial check on the various dualities 
we have used.
\\
Now, the $p_1$ term for the $D3$ has already been computed
\cite{das,bac} (by S-duality):
\begin{equation}
 f_{D3}(\tau)= \frac{1}{4 \pi}Re \left[ \, i\, \ln(\eta(\tau)) \right]
\label{d}
\end{equation}
So, we solve the independent equations \eqref{a} and get
\begin{eqnarray}
f_{O3^{-}}(\tau)&=&\frac{1}{32 \pi}Re \left[ \, i\, \ln(\eta(\tau)) \right]
\label{e} \\
f_{O3^{+}}(\tau)&=&-\frac{1}{32 \pi}Re \left[ \, i\, \ln\left(\frac{\eta(\tau)}{\vartheta_{2}^{2}(\tau)}\right) \right]
\label{f}
\end{eqnarray}
By applying  a transformation $S$ or  $ST$ on $O3^{+}$, we obtain
respectively the couplings for the 
$\widetilde{O3^{-}}$ and $\widetilde{O3^{+}}$ planes:
\begin{eqnarray}
f_{\widetilde{O3^{-}}}(\tau)&=&-\frac{1}{32
\pi}Re \left[ \, i\, \ln\left(\frac{\eta(\tau)}{\vartheta_{4}^{2}(\tau)}\right) \right]
\label{g}
\\
f_{\widetilde{O3^{+}}}(\tau)&=&-\frac{1}{32 \pi}Re \left[ \, i\, \ln\left(\frac{\eta(\tau)}{\vartheta_{3}^{2}(\tau)}\right) \right]
\label{i}
\end{eqnarray}
The expressions \eqref{e}, \eqref{f} in the weak coupling regime $e^{-\phi}
\rightarrow \infty$ reproduce
the known tree level coupling $p_1$. The exponential terms $\sum_{n}\mu(n)e^{2i\pi n\tau}$ are identified with
the $D_{-1}$ contributions.
The expressions \eqref{e}, \eqref{f}, \eqref{g}, \eqref{i}  are also invariant under the right duality groups.
Indeed, the groups $SL(2,{ \mathbb{Z}})$, $\Gamma^{0}(2)$, $\Gamma_{0}(2)$ and $\Gamma(2)$ correspond
to the invariance groups (modulo phases and $\, \ln\left({\tau \over \bar{\tau}}
\right)$ terms) of $\eta$, $\vartheta_{4}$, $\vartheta_{2}$, $\vartheta_{3}$ respectively.

Moreover, expanding $f_{\widetilde{O3^{-}}}(\tau)$ and  $f_{\widetilde{O3^{+}}}(\tau)$
in the weak coupling regime, we find
\begin{eqnarray}
 f_{\widetilde{O3^{-}}}(\tau)&=&\frac{1}{384}C_{0}+O(e^{2i\pi n\tau}) \\
 f_{\widetilde{O3^{+}}}(\tau)&=&\frac{1}{384}C_{0}+O(e^{2i\pi
n\tau}) \end{eqnarray}
By applying T-duality on $\widetilde{O3^{-}}$ and
$\widetilde{O3^{+}}$, we obtain the Wess-Zumino terms for all $p \leq
5$
\begin{eqnarray}
  WZ \left(\widetilde{Op^{-}}\right)&=&(-2^{p-5}+{1 \over
2})\int_{M_{p+1}} C \wedge \sqrt{\hat{L}(R/4)} 
\label{Zi}
\\
 WZ \left(\widetilde{Op^{+}}\right)&=&\int_{M_{p+1}} C \wedge
 \left((-2^{p-5}+{1 \over 2})\sqrt{\hat{L}(R/4)} +\sqrt{\hat{A}(R)}(2^{p-4}-
{1 \over 2})\right)
 \label{Xi}
\end{eqnarray}
We could define the Wess-Zumino action for a half $Dp$ brane as the
difference between $WZ(\widetilde{Op^{-}})$ and
$WZ(\widetilde{Op})$. We obtain for $p\leq
5$
 \begin{eqnarray}
  WZ \left({Dp \over 2}\right)&=&{1 \over
2}\int_{M_{p+1}} C \wedge \sqrt{\hat{L}(R/4)} 
\end{eqnarray}    

These results differ from \cite{muk} and this can be traced back to
the fact that $WZ(\widetilde{O4^{-}})$ is different from zero in there.
Our results will be 
reproduced in the next
section through computations in ${\cal M}$-theory.
Note also that the difference
\eqref{Xi}-\eqref{Zi} depends only on
the square root of the roof genus as was the case for the difference
\eqref{Op^{-}}-\eqref{Op^{+}}.

\section{${\cal M}$-theory interpretation of Orientifold Planes}

We intend to compare these results with graviton scattering in eleven
dimensional supergravity at one-loop. But, first, let us  review the 
${\cal M}$-theory interpretation of orientifold four planes
\cite{hor,gim,han}. We shall denote by $S_n$ a circle along the
n$^{\text{th}}$ direction. 
\\
a) $O4^{-}$ is lifted to ${\cal M}$-theory
on $\mathbb{R}^{4,1} \times {\mathbb{R}^5 \over \mathbb{Z}_2} \times S_{11}$ with
$\mathbb{Z}_2$ the $\mathbb{R}^5$ parity. The fixed plane ${\mathbb{R}^5 \over \mathbb{Z}_2}$ called ${\cal
OM}5$ must carry $-{1 \over 2}$ the unit charge of an $M5$ brane in order to
cancel the $6d$ gravitational anomaly \cite{wit3}. We do not need to add a twisted sector on this plane. Thus,
this object carries the same charge as the $O4^{-}$ ( $Q_4=-{1
\over 2}$).
Compactifying
$O4^{-}$ on $S_{9}$ and applying a
 T-duality along this direction, we obtain two $O3^{-}$'s in type
$IIB$.  Now this configuration  is lifted to 
$\mathbb{R}^{3,1} \times {\mathbb{R}^5 \over \mathbb{Z}_2} \times T^2$ with
the two-torus $T^2$ along the nine and eleven directions.
The complex $IIB$ coupling $\tau$ is identified with the complex
structure $U={R_{9} \over R_{11}}e^{i\theta} $ of the two-torus $T^2$ where
$\theta$ is the angle of the torus whose cycles have length $2\pi R_9$
and $2\pi R_{11}$.
Thus, according to \eqref{e}, ${\cal OM}5$  wrapped on $T^2$ should have a term $\int_{\mathbb{R}^{3,1}}{1 \over 16 
\pi}Re \left[ \, i\, \ln(\eta(\tau)) \right]p_1$.
\\
b) $O4^{+}$  is lifted to ${\cal M}$-theory
on $\mathbb{R}^{4,1} \times {\mathbb{R}^5 \over \mathbb{Z}_2} \times S_{11}$  with a
$M5$ brane.
This $M5$ is 
stuck on the fixed plane of $\mathbb{Z}_2$ by imposing as holonomy
along $S_{11}$ an
element of $O(2)$, $W=\left(
\begin{array}{ll}
1
& 0 \cr
0 & -1 \cr
\end{array}
\right)
$  which is not connected to the identity \cite{gim}.

This configuration reproduces the charges of the $O4^{+}$.
Compactifying
$O4^{+}$ on $S_{9}$ and applying a T-duality, we obtain that a $IIB$
configuration with 2 
$O3^{+}$ is lifted to an ${\cal
OM}5$ and a $M5$ brane both stuck at the fixed point of $\mathbb{Z}_2$ 
and wrapped
on $T^2$ with a Wilson line $W$.
We can deduce from \eqref{f}-\eqref{e} 
 that the contribution of the $M5$ alone must be equal to the difference 
 \begin{equation}
2WZ(O3^+-O3^-) = \int_{\mathbb{R}^{3,1}}{1 \over 4 
\pi}Re \left[\, i\, \ln\left({\eta(2\tau) \over \eta(\tau)}\right) \right]p_1
\label{r1}
\end{equation}
\\
c) $\widetilde{O4^{-}}$ is lifted to ${\cal M}$-theory
on  $\mathbb{R}^{4,1} \times {\mathbb{R}^5 \times S_{11} \over \mathbb{Z}_2}$
 with
 $\mathbb{Z}_2$ now the  $\mathbb{R}^5$ parity times the action 
on the circle $S_{11}$ by a shift: $x_{11} \rightarrow x_{11}+\pi$.
As there is no fixed point, there is no ${\cal
OM}5$ and no $M5$. Moreover, the shift produces a non-trivial
``discrete torsion'' for the RR 1-form $C_{1}$ in type $IIA$
\cite{gim,hor}.
 So, this
configuration reproduces the fact that $\widetilde{O4^{-}}$ has a
vanishing $Q_4$ charge and a non-trivial RR flux for $\C_{1}$.
 Compactifying on $S_{9}$ and applying a
T-duality, we obtain one $O3^{-}$ and one $\widetilde{O3^{-}}$. So, {\cal
M}-theory on $\mathbb{R}^{3,1} \times {\mathbb{R}^5 \times S_{11} \over \mathbb{Z}_2} \times S_{9}$ should have  a
term ${1 \over 16 
\pi}Re \left[ \, i\, \ln(\vartheta_{4}(\tau)) \right]p_1$ according to  \eqref{e} and \eqref{g}.
\\
d) Finally, $\widetilde{O4^{+}}$ wrapped on $S_{9}$ (which has also a non trivial discrete
torsion for the 1-form RR in type $IIA$) is lifted to ${\cal M}$-theory
on  $\mathbb{R}^{3,1} \times {\mathbb{R}^5 \times S_{11} \over \mathbb{Z}_2} \times S_{9}$ with a
$M5$ wrapped on $S_{9}$ and stuck at the origin on the circle of half radius
$S_{11} \over \mathbb{Z}_2$. Applying a
T-duality along $S_{9}$, we obtain that a pair of  
$\widetilde{O3^{+}}$ and $O3^{+}$ orientifolds are lifted 
to ${\cal M}$-theory
on  $\mathbb{R}^{3,1} \times {R^5 \times T^2_{9 11} \over \mathbb{Z}_2}$ with a
$M5$.
To be compatible with our previous results, the $M5$, in this new
geometry,  
must have a term
\begin{equation} 
WZ(\widetilde{O3^{+}}+O3^{+}-\widetilde{O3^{-}}-O3^{-})=
\int_{\mathbb{R}^{3,1}} {1 \over 4 
\pi}Re \left[ \, i\, \ln\left({\eta(\tau) \over \eta({\tau \over 2})}\right) \right]p_1
\label{r2}
\end{equation}
\bigskip

Now we would like to reproduce the terms \eqref{r1}, \eqref{r2} 
which are the ${\cal M}$-theory consequences of
\eqref{d}, \eqref{e}, \eqref{f}, \eqref{g}, \eqref{i} 
from a one- loop computation 
in ${\cal M}$-theory.
The topological origin 
in ${\cal M}$-theory of gravitational couplings $\int C_1p_1$ for four planes
 is not clear. It was argued in 
\cite{muk1} that the chirality of $M5$ and ${\cal OM}5$ gives the right
Wess-Zumino actions for the $D4$ brane, orientifolds $O4^-$ and $O4^+$,
after compactifying on a circle. As the orientifold
$\widetilde{O4^{-}}$ is not associated to any $M5$ or ${\cal OM}5$ 
in ${\cal M}$-theory,
it is quite encouraging to find that its Wess-Zumino action vanishes.
Although we do not know how to reproduce the terms \eqref{r1}, \eqref{r2}
for the $M5$ brane, we are going to reproduce  terms
 proportional to $trR\wedge \star
R$. As we have at least ${\cal N}=1$ in $d=4$,  
the CP-even and  odd parts $A_{even}$ and $A_{odd}$ are the
same up to a factor $i$ \cite{ant}.
 
A M5 wrapped on a two torus $T^2$ with a Wilson line $W$ must
have
 a term 
\begin{equation}
 \int_{\mathbb{R}^{3,1}}{1 \over 4 
\pi}Re \left[ \, \, \ln\left({\eta(\tau) \over \eta({2\tau})}\right) \right]trR\wedge \star
R
\label{r3}
\end{equation}
Moreover a M5 wrapped on $S_{9}$ and stuck on a circle of half radius
reproduces the term
\begin{equation}
 \int_{\mathbb{R}^{3,1}}{1 \over 4 
\pi}Re \left[ \, \, \ln\left({\eta({\tau \over 2}) \over \eta(\tau)}\right) \right]trR\wedge \star
R
\label{r4}
\end{equation}

\begin{figure}[http]
\centering
\epsfxsize=2in
\epsfysize=2in
\epsffile{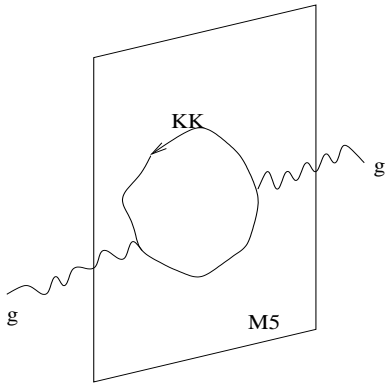}
\caption{ Two external gravitons
scattering in the world volume of a $M5$ brane. }
\label{diagr}
\end{figure}

Let us now consider a one-loop diagram with two external gravitons
scattering in the world volume of a $M5$ brane (Fig. \ref{diagr}).

This one-loop effect can be computed as the sum of Feynman diagrams
with the (2,0) tensor multiplet of the $M5$ circulating in the loop.
The Kaluza-Klein modes with momenta $p_{11}={ l_{11} \over R_{11}}$
and $p_{9}={ l_{9} \over R_{9}}$ 
circulating in the loop will be interpreted as a bound state of
$l_{11}$ D0 branes that moves along an euclidean times $x^{9}$ in type
$IIA$  and as all $(l_9,l_{11})$ D-instantons
in $IIB$ theory on the orientifold plane.
This loop amplitude can be also calculated by using the light
cone description of the superparticle \cite{gre1}. 
   
The expression for the one-loop amplitude is \cite{gre1}, \cite{bac}:

\begin{equation}
A_{even}=\int_{\mathbb{R}^{3,1}}{1 \over 16 \pi}trR \wedge \star
R \int_{0}^{\infty}{dt \over
t}\sum_{l_I}{e^{-\pi tG^{IJ}l_Il_J}}
\label{A}
\end{equation}
\noindent
where $G^{IJ}l_Il_J=
\frac{|l_9-l_{11}\tau|^2}{V_{2}\tau_2}$.
The overall normalization is chosen in order to reproduce \eqref{d} 
the  correction for a D3 brane \cite{bac}. A T-dualized D3 is lifted
 to a M5 wrapped on a two torus, in this case  
$l_I \in \mathbb{Z}^2$. Finally
\begin{equation}
A_{even}= \int_{\mathbb{R}^{3,1}}{1 \over 16 \pi}trR \wedge \star
R \int_{0}^{\infty}{dt \over
t}\sum_{l_I \in \mathbb{Z}^2 }{e^{-\pi tG^{IJ}l_Il_J}}
\end{equation}
Doing a Poisson resummation, we obtain
\begin{equation}
A_{even}= \int_{\mathbb{R}^{3,1}}{1 \over 16 \pi^2}trR \wedge \star
R \sum_{w^I \in \mathbb{Z}^2 }\frac{\tau_2}{|w_{11}-w_9\tau|^2}
\end{equation}
This Maass function can be regularized by eliminating the pole at $s=1$:
\begin{equation}
\hat{E}_1=\lim_{s \rightarrow 1}\left(\sum_{w^I \in \mathbb{Z}^2
\setminus (0,0)}\left(\frac{\tau_2}{|w_{11}-w_9\tau|^2}\right)^s-\frac{\pi}{s-1}\right)=-4\pi
Re \left[\, \ln\left(\eta(\tau)\right) \right]
\label{reg}
\end{equation}
modulo a perturbatively invisible logarithmic term   $-\pi
\, \ln(\tau_2)$, not present in the Wilsonian effective action of the
orientifolds and D3 branes, so we have removed it by hand.
So, we obtain \begin{equation}
A_{even}=-{1 \over 4 \pi}Re \left[\, \ln\left(\eta(\tau)\right) \right] \int_{\mathbb{R}^{3,1}}trR \wedge \star
R 
\end{equation}
\noindent
Now, we need to determine properly the Kaluza-Klein modes which are
identified with $(l_9,l_{11})$ D-instantons, ie the
ranges of values of $(l_9,l_{11})$ for the other configurations.
We use that the number of bound states for $(0,1)$
instantons on an orientifold $O3^-$ or $\widetilde{O3^-}$ (resp. 
$O3^+$ or  $\widetilde{O3^+}$) is integer (resp. half integer) as the
gauge group is $SO$ (resp. $Sp$). This analysis has been done for the
system $(0,1)$ instanton (ie $D_{-1}$) on $O3^{\pm}$ in \cite{pol} and
for the system $D_{-1}$ on $\widetilde{O3^\pm}$ in \cite{wit2}.
This can also be understood in the picture $IIA$-${\cal M}$ \cite{gim}.
Now, doing a S-duality exchanges $O3^+$ with $\widetilde{O3^-}$
and permutes $l_9$ with $l_{11}$. A transformation T exchanges
$\widetilde{O3^-}$ with $\widetilde{O3^+}$ and shifts 
$l_{9}$ by $l_{11}$. So, using the various self-duality subgroups preserving
 the orientifold 3-planes (see Table 1), we list the number of 
D-instantons living on them the solution in Table 2 below.
\begin{table}[http]
\begin{center}
\begin{tabular}{|c|c|c|c|c|}
\hline
$(l_9,l_{11})$ mod 2  & $O3^-$ & $O3^+$ & $\widetilde{O3^-}$ & $\widetilde{O3^+}$
\\
\hline
 $(0,1)$ & $1$ & ${1 \over 2}$ & $1$ 
 & ${1 \over 2}$   
\\
\hline
 $(1,0)$ & $1$ & $1$ & ${1 \over 2}$ 
 & ${1 \over 2}$  
\\
\hline
 $(1,1)$ & $1$ & ${1 \over 2}$ & ${1 \over 2}$ 
 & $1$
\\  
\hline
 
\end{tabular}
\end{center}

\label{inst}
\caption{Minimal  topological charges for $(l_9,l_{11})$ instantons on orientifold three-planes.}

\end{table}
\\
   
Now let us compute in the two cases including an $M5$ brane. We first want to 
reproduce the expression for a $M5$ 
wrapped on $T^2$ with a Wilson line
W. The Kaluza-Klein circulating in
the loop correspond to instantons stuck on  $O3^+$ and we need to
sum the one-loop amplitude over ${\mathbb{Z}  \over 2}. (0,1)$ , 
 $\mathbb{Z}. (1,0)$ and  ${\mathbb{Z} \over 2}. (1,0)$. 

First, the one-loop amplitude is a sum over ${\mathbb{Z}  \over
2}. (0,1)$, this 
means that we must sum \eqref{A} over $( l_9 \in \mathbb{Z}, l_{11} \in 
\mathbb{Z}+{1 \over 2})$ and  $( 2\mathbb{Z}, 
2\mathbb{Z}+1)$:

\begin{equation}
A_{even}^{{\mathbb{Z}  \over 2}. (0,1)}={1 \over 16 \pi}\int_{\mathbb{R}^{3,1}}trR \wedge \star
R \int_{0}^{\infty}{dt \over
t}\left(\sum_{(l_9 \in \mathbb{Z},l_{11} \in 
\mathbb{Z}+{1 \over 2})}+\sum_{( 2\mathbb{Z}, 
2\mathbb{Z}+1)}\right) {e^{-\pi tG^{IJ}l_Il_J}}
\end{equation} 

By rescaling the second summation, we obtain
\begin{equation}
A_{even}^{{\mathbb{Z}  \over 2}. (0,1)}=2{1 \over 16 \pi}\int_{\mathbb{R}^{3,1}}trR \wedge \star
R \int_{0}^{\infty}{dt \over
t}\sum_{(l_9 \in \mathbb{Z},l_{11} \in 
\mathbb{Z}+{1 \over 2})}{e^{-\pi tG^{IJ}l_Il_J}}
\end{equation} 
This can be written as:
\begin{equation}
A_{even}^{{\mathbb{Z}  \over 2}. (0,1)}=2{1 \over 16 \pi}\int_{\mathbb{R}^{3,1}}trR \wedge \star
R \int_{0}^{\infty}{dt \over
t}\left(\sum_{(l_9 \in \mathbb{Z},l_{11} \in 
\frac{\mathbb{Z}}{2})}-\sum_{(\mathbb{Z}, 
\mathbb{Z})}\right){e^{-\pi tG^{IJ}l_Il_J}}
\end{equation} 
Using the regularization \eqref{reg}, we obtain the result
\begin{equation}
A_{even}^{{\mathbb{Z}  \over 2}. (0,1)}= \int_{\mathbb{R}^{3,1}}{1 \over 4 
\pi}Re \left[ \, \, \ln\left({\eta(\tau) \over \theta_{4}(\tau)}\right) \right]trR\wedge \star
R
\label{r6}
\end{equation}
By analogy with the previous rule, the one-loop amplitude over 
$\mathbb{Z}. (1,0)$ is 
\begin{equation}
A_{even}^{\mathbb{Z} . (1,0)}={1 \over 16 \pi}\int_{\mathbb{R}^{3,1}}trR \wedge \star
R \int_{0}^{\infty}{dt \over
t}\sum_{(l_9 \in 2\mathbb{Z}+1,l_{11} \in 
2\mathbb{Z})}{e^{-\pi tG^{IJ}l_Il_J}}
\end{equation} 
Doing the same kind of algebraic manipulation as previously, we obtain
\begin{equation}
 A_{even}^{\mathbb{Z} . (1,0)}=\int_{\mathbb{R}^{3,1}}{1 \over 8 
\pi}Re \left[ \, \, \ln\left({\eta(\tau) \over \theta_{2}(\tau)}\right) \right]trR\wedge \star
R
\label{r7}
\end{equation}
Moreover, by applying some $SL(2,\mathbb{Z})$ transformations, 
we can obtain the
following one- loop amplitudes:

\begin{eqnarray}
  A_{even}^{{\mathbb{Z} \over 2}.(1,1)}&=&\int_{\mathbb{R}^{3,1}}{1 \over 4 
\pi}Re \left[ \, \, \ln\left({\eta(\tau) \over
\theta_{3}(\tau)}\right) \right]trR\wedge \star R 
\label{r8}
\\
 A_{even}^{{\mathbb{Z} \over 2} . (1,0)}&=&\int_{\mathbb{R}^{3,1}}{1 \over 8 
\pi}Re \left[ \, \, \ln\left({\eta(\tau) \over
\theta_{2}(\tau)}\right) \right]trR\wedge \star R 
\label{r9}
\\
 A_{even}^{\mathbb{Z}  . (1,1)}&=&\int_{\mathbb{R}^{3,1}}{1 \over 8 
\pi}Re \left[ \, \, \ln\left({\eta(\tau) \over
\theta_{3}(\tau)}\right) \right]trR\wedge \star R
\label{r10}
\end{eqnarray}

Finally, the summation over the D-instantons stuck on $O3^+$ is 
\begin{equation}
A_{even}^{{\mathbb{Z} \over 2}  . (0,1)}+
A_{even}^{\mathbb{Z}   . (1,0)}+A_{even}^{{\mathbb{Z} \over 2}
. (1,1)}=\int_{\mathbb{R}^{3,1}}{1 \over 4 
\pi}Re \left[ \, \, \ln\left({\eta(\tau) \over
\eta(2\tau)}\right) \right]trR\wedge \star R
\end{equation}
We obtain the right result \eqref{r3}.
\\
Now, we do the same thing for $M5$ wrapped on $S_9$ and stuck on a
circle of half radius $S_{11}/2$. We have seen previously that the
Kaluza-Klein modes stuck on $M5$ correspond to D-instantons on $O3^+$
and $\widetilde{O3^-}$. So, the one-loop amplitude , according to the
table 2, is:

\begin{equation}
2A_{even}^{{\mathbb{Z} \over 2}  . (0,1)}+
A_{even}^{\mathbb{Z}   . (1,0)}+A_{even}^{{\mathbb{Z} \over 2}
. (1,1)}+A_{even}^{{\mathbb{Z} \over 2}
. (1,0)}+A_{even}^{\mathbb{Z}   . (1,1)}=\int_{\mathbb{R}^{3,1}}{1 \over 4 
\pi}Re \left[ \, \, \ln\left({\eta({\tau \over 2}) \over
\eta(\tau)}\right) \right]trR\wedge \star R
\end{equation}
\noindent
We obtain the right result \eqref{r4}

\section{Conclusion:}
Using various string dualities, we have found the Wess-Zumino actions
for orientifold planes with a non trivial $RR$ flux. The $p=7$ analog
problem remains open. The precise rules of instantonic computation are
not known but this could be circumvented by using S, T dualities. We
have used (following \cite{gim}) the lift of orientifolds
$\widetilde{Op^{
\pm}}$ in ${\cal M}$ theory  without giving a name as they are not
localized. But we found a set of rules for D-instantons living on
orientifold three planes.

\vskip 0.3cm
\noindent{\bf \large Acknowledgments}
\vskip 0.2cm
\noindent
We have benefited from
discussions with C. Bachas, P. Bain, A. Hanany, 
E. Kiritsis, L. Paulot, P. Vanhove and especially A. Keurentjes.

\vskip 1cm
\noindent{\bf \large Appendix A: Properties of Theta functions }
\vskip 0.1cm
\noindent
we set $q=e^{2i\pi\tau}$
\begin{eqnarray}
\eta(\tau)&=&q^{1 \over 24}\prod_{n=1}^{\infty}{(1-q^n)} 
\\
\vartheta_2(\tau)&=&2q^{1 \over
8}\prod_{n=1}^{\infty}{(1-q^n)(1+q^{n})^2}=2\frac{\eta^2(2\tau)}{\eta(\tau)} 
\\
\vartheta_3(\tau)&=&\prod_{n=1}^{\infty}{(1-q^n)(1+q^{n+{1 \over
2}})^2}=2{e^{-i\pi \over 3} \over  1-\tau}\frac{\eta^2(\frac{1+\tau}{1-\tau})}{\eta(\tau)}
\\
\vartheta_4(\tau)&=&\prod_{n=1}^{\infty}{(1-q^n)(1-q^{n-{1 \over 2}})^2}
=\frac{\eta^2({\tau \over 2})}{\eta(\tau)}
\\
2\eta^3(\tau)&=&\vartheta_2 \vartheta_3 \vartheta_4
\end{eqnarray}

\begin{table}[http]
\begin{center}
\begin{tabular}{|c|c|c|}
\hline
  & S & T 
\\
\hline
$\eta(\tau)$ & $\sqrt{-i\tau}\eta(\tau)$ & $e^{\frac{i\pi}{12}}\eta(\tau)$
\\
\hline
$\vartheta_2(\tau)$  & $\sqrt{-i\tau}\vartheta_4(\tau)$ & $e^{i\pi \over 4}\vartheta_2(\tau)$ 
\\
\hline
$\vartheta_3(\tau)$  & $\sqrt{-i\tau}\vartheta_3(\tau)$ & $\vartheta_4(\tau)$
\\
\hline
$\vartheta_4(\tau)$  & $\sqrt{-i\tau}\vartheta_2(\tau)$ & $\vartheta_3(\tau)$
\\
\hline
\end{tabular}
\end{center}

\label{theta}
\caption{
Transformations of the Theta functions under the generators S
and T of $SL(2,\mathbb{Z})$}

\end{table}

Poisson's Resummation:

\begin{equation}
\sum_{\vec{l} \in
Z}
e^{-\pi(\vec{l}+\vec{x}).A.(\vec{l}+\vec{x})}
=A^{-{1 \over 2}}\sum_{\vec{w} \in
Z}e^{-\pi\vec{w}.A^{-1}.\vec{w}+2\pi i\vec{w}.\vec{x}}
\end{equation}

\vskip 1cm
\noindent{\bf \large Appendix B: The puzzle of seven planes }
\vskip 0.1cm
\noindent

The Green-Schwarz terms for the Heterotic string compactified on $T^2$
(with gauge group $SO(8)^4$), dual to IIB orientifold with 16 $D7$ and 4
$O7^-$, has already been computed in \cite{kir}. 
For example, we can deduce that 
\begin{equation}
\int \left(16f_{D7}(\tau)+4f_{O7^{-}}(\tau)\right)p_2=
\int \frac{1}{120\pi}Re \left[i\, \ln(\vartheta_4(\tau)\eta(\tau)) \right]p_2
\label{gg}
\end{equation}
Now, we will assume that the Wess-Zumino action of $O7^-$ can be obtained by
adding the action of  a (1,1) and (1,-1) 7-planes. 
We can note that even though these (p,q) 7-planes are associated with strong coupling
and may not have a well-defined Lagrangian description,
the Wess-Zumino terms are topological and must make sense at least for
anomaly cancelling reasons.

By definition, the
action of a 
(p,q) 7-plane is found by applying the inverse of a   
  $g_{(p,q)}=
\left(
\begin{array}{ll}
p
& r  \cr
q & s  \cr
\end{array}
\right)$ $\epsilon$ $SL(2,{ \mathbb{Z}})$ transformation on the action
of a $D7$ brane.
So, the function $f_{D7}(\tau)$ must be invariant under T so that the
action for the (p,q) branes does not
depend on the integers r and s. 
Then, if we ignore possible interaction terms, we obtain:
\begin{equation}
16f_{D7}(\tau)+4\left(f_{D7}(g^{-1}_{(1,1)}\tau)+f_{D7}(g^{-1}_{(1,-1)}\tau)\right)=
\frac{1}{120\pi}Re \left[i\, \ln(\vartheta_4(\tau)\eta(\tau)) \right]
\label{r}
\end{equation}
Algebraic manipulations imply that T-invariance of $D7$ cannot hold.

Thus,
some work remains to be done in order to find the non-perturbative
D-instanton corrections for $D7$, $O7^-$ and $O7^+$.


\begin{thebibliography}{99}

\bibitem{wit1}
E. Witten, {\sl Baryons And Branes In Anti de Sitter Space}, JHEP
9807 (1998) 006, hep-th/9805112

\bibitem{han}
A. Hanany, B. Kol, {\sl On Orientifolds, Discrete Torsion,
Branes and M Theory}, JHEP 0006 (2000) 013, hep-th/0003025

\bibitem{cra}
B. Craps, F. Roose, {\sl Anomalous D-brane and orientifold
couplings from the boundary state}, Phys.Lett. B445 (1998) 150-159, 
hep-th/9808074

\bibitem{mor}
J.F. Morales, C.A. Scrucca, M. Serone, {\sl Anomalous couplings for
D-branes and O-planes}, Nucl. Phys. B552 (1999) 291-315,
hep-th/9812071


\bibitem{ste}
B. Stefanski Jr,
{\sl Gravitational Couplings of D-branes and O-planes}, 
Nucl. Phys. B548 (1999) 275-290, hep-th/9812088

\bibitem{gre}
M. Green, J. A. Harvey, G. Moore, {\sl I-Brane Inflow
and Anomalous Couplings on D-Branes}, Class.Quant.Grav. 14 (1997)
47-52, hep-th/9605033

\bibitem{scr}
C.A. Scrucca, M. Serone,
{\sl Anomalies and inflow on D-branes and O-planes}, Nucl. Phys. B556
(1999) 197-221, hep-th/9903145


\bibitem{muk}
S. Mukhi, N. V. Suryanarayana,
{\sl Gravitational Couplings, Orientifolds and M-Planes}, JHEP 9909
(1999) 017, hep-th/9907215

\bibitem{keu}
A. Keurentjes, {\sl Classifying orientifolds by flat n-gerbes},
 hep-th/0106267

\bibitem{gim}
E. G. Gimon, {\sl On the M-theory Interpretation of Orientifold
Planes}, hep-th/9806226

\bibitem{sen}
A. Sen, {\sl F-theory and Orientifolds}, Nucl.Phys. B475 (1996)
562-578, hep-th/9605150

\bibitem{sch}
J. H. Schwarz, {\sl An SL(2,$\mathbb{Z}$) Multiplet of Type IIB Superstrings},
Phys.Lett. B360 (1995) 13-18; Erratum-ibid. B364 (1995) 252, hep-th/9508143

\bibitem{wit2}
E. Witten, {\sl Toroidal Compactification Without Vector
Structure}, JHEP 9802 (1998) 006, hep-th/9712028

\bibitem{bia}
M. Bianchi, {\sl A Note on Toroidal Compactifications of the Type I 
Superstring and Other Superstring Vacuum Configurations with 16 Supercharges}
Nucl.Phys. B528 (1998) 73-94, hep-th/9711201

\bibitem{cha}
S. Chaudhuri, G. Hockney, J. D. Lykken, {\sl Maximally Supersymmetric
String Theories in $D \le 10$}, Phys.Rev.Lett. 75 (1995)
2264-2267, hep-th/9505054

\bibitem{lsch}
W. Lerche, R. Minasian, C. Schweigert, S. Theisen, {\sl A Note on the 
Geometry of CHL Heterotic Strings}, Phys.Lett. B424 (1998) 53-59, 
hep-th/9711104

\bibitem{greg}
A. Gregori, E. Kiritsis, C. Kounnas, N. A. Obers, P. M. Petropoulos,
B.Pioline, {\sl $R^2$ Corrections and Non-perturbative Dualities of $N=4$
String ground states}, Nucl.Phys. B510 (1998) 423-476, hep-th/9708062

\bibitem{vafa}
A. Sen, C. Vafa, {\sl Dual Pairs of Type II String Compactification
}, Nucl.Phys. B455 (1995) 165, hep-th/9508064
 
\bibitem{das}
K. Dasgupta, D. P. Jatkar, S. Mukhi, {\sl Gravitational Couplings and ${ \mathbb{Z}}_2$ Orientifolds},
Nucl.Phys. B523 (1998) 465-484, hep-th/9707224



\bibitem{bac}
C. P. Bachas, P. Bain, M. B. Green, {\sl Curvature terms in D-brane
actions and their M-theory origin }, JHEP 9905 (1999) 011,
hep-th/9903210

\bibitem{hor}
K. Hori, {\sl Consistency Conditions for Fivebrane in M Theory on $R^5/\mathbb{Z}_2$
Orbifold}, Nucl.Phys. B539 (1999) 35-78, hep-th/9805141


 

\bibitem{wit3}
E. Witten, {\sl Five-branes And $M$-Theory On An Orbifold},
Nucl.Phys. B463 (1996) 383-397, hep-th/9512219

\bibitem{muk1}
S. Mukhi, {\sl Dualities and the SL(2,Z) Anomaly}, JHEP 9812 (1998)
006, hep-th/9810213

\bibitem{ant}
I. Antoniadis, E. Gava, T. R. Taylor, Phys.Lett.B 267 (1991), 37 

\bibitem{gre1}
M.B. Green, M. Gutperle, P. Vanhove, {\sl One loop in eleven
dimensions}, Phys.Lett. B409 (1997) 177-184, hep-th/9706175

\bibitem{pol}
E. G. Gimon, J. Polchinski, {\sl Consistency Conditions for
Orientifolds and D-Manifolds}, Phys.Rev. D54 (1996) 1667-1676, 
hep-th/9601038

\bibitem{wit2}
Edward Witten, {\sl Toroidal Compactification Without Vector
Structure}, HEP 9802 (1998) 006, hep-th/9712028


\bibitem{kir}
E. Kiritsis, N. A. Obers, B. Pioline,
{\sl Heterotic / Type II Triality and Instantons on K(3)},JHEP 0001:029,2000, hep-th/0001083


\bibitem{ler}
W. Lerche, {\sl On the Heterotic/F-Theory Duality in Eight Dimensions},
 proceedings of Cargese 1999, hep-th/9910207




\end{thebibliography}
\end{document}